\documentclass[prd,amstext,amsmath,amssymb,superscriptaddress,floatfix,nofootinbib,notitlepage]{revtex4-1}

\usepackage[hyperref]{xcolor}
\usepackage{graphics}
\usepackage{amsmath}
\usepackage{amssymb}
\usepackage{amsthm}
\usepackage{graphicx}
\usepackage{subfigure}
\usepackage{slashed}
\usepackage{hyperref}
\hypersetup{
  colorlinks, breaklinks,
  linkcolor=blue,
  citecolor=blue,
  linktoc=all
}

\definecolor{stcol}{rgb}{1,0,1}
\definecolor{mcol}{rgb}{0,0,1}

\usepackage{ulem} 

\newcommand{\Eq}[1]{{Eq.~({\ref{#1}})}}
\newcommand{\Eqs}[1]{{Eqs.~({\ref{#1}})}}
\newcommand{\Fig}[1]{{Fig.~{\ref{#1}}}}
\newcommand{\ave}[1]{{\langle{#1}\rangle}}
\newcommand{\bea}{\begin{eqnarray}}
\newcommand{\eea}{\end{eqnarray}}
\newcommand{\beq}{\begin{equation}}
\newcommand{\eeq}{\end{equation}}
\newcommand{\beas}{\begin{eqnarray*}}
\newcommand{\eeas}{\end{eqnarray*}}

\def\p{{\bf p}}
\def\q{{\bf q}}
\def\x{{\bf x}}

\graphicspath{
{./}
}

\begin{document}

\title{Inhomogeneous chiral phases away from the chiral limit}

\author{Michael~Buballa}
\affiliation{Theoriezentrum, Institut f\"ur Kernphysik, Technische Universit\"at Darmstadt, 
Schlossgartenstr.\ 2, D-64289 Darmstadt, Germany}

\author{Stefano~Carignano}
\affiliation{Departament de F\'isica Qu\`antica i Astrof\'isica and Institut de Ci\`encies del Cosmos, Universitat de Barcelona, Mart\'i i Franqu\`es 1, 08028 Barcelona, Catalonia, Spain.}

\begin{abstract}

The effect of explicit chiral-symmetry breaking on inhomogeneous chiral phases is studied within a 
Nambu--Jona-Lasinio model with nonzero current quark mass.
Generalizing an earlier result obtained in the chiral limit, we show within a Ginzburg-Landau analysis
that the critical endpoint of the first-order chiral phase boundary between two homogeneous phases 
gets replaced by a ``pseudo-Lifshitz point'' when the possibility of inhomogeneous order parameters
is considered. 
Performing a stability analysis we also show that the unstable mode along the phase boundary 
is in the scalar but not in the pseudoscalar channel, suggesting that modulations which contain pseudoscalar 
condensates, like a generalized dual chiral density wave, are disfavored against purely scalar ones.
Numerically we find that the inhomogeneous phase shrinks as one moves away from the chiral limit, but survives 
even at significantly large values of the current quark mass.

\end{abstract}

\maketitle

\section{Introduction}

The conjectured existence of a chiral critical point in the QCD phase diagram at nonvanishing temperature $T$ and chemical
potential $\mu$~\cite{Stephanov:1998dy} has triggered tremendous experimental and theoretical activities, e.g.,
\cite{Stephanov:1999zu,Fodor:2004nz,Friman:2011zz,Kumar:2013cqa}. 
While lattice gauge simulations with physical quark masses have revealed that at $\mu=0$ the transition
from the low-temperature phase with spontaneously broken chiral symmetry to the approximately restored phase at high
temperature is not a true phase transition but only a smooth crossover~\cite{Aoki:2006we}, 
the situation is still unsettled at nonvanishing chemical potential where the application of standard lattice techniques is 
hampered by the sign problem.
In this regime, calculations within effective models, like the Nambu--Jona-Lasinio (NJL) model or the quark-meson (QM) model,
typically predict the existence of a first-order phase boundary, where the chiral order parameter changes discontinuously. 
When increasing the temperature, the discontinuity decreases until the first-order phase boundary vanishes at a critical endpoint 
(CEP) \cite{Asakawa:1989bq,Scavenius:2000qd}.
Beyond the CEP one finds again a crossover, consistent with the 
lattice results at $\mu = 0$. A special case is the chiral limit, where an exactly chirally restored phase exists.
For two quark flavors, instead of a crossover, one then finds a second-order phase transition at high temperature  
and instead of a CEP there is a tricritical point (TCP) where the second-order phase boundary joins with the first-order one.

A tacit assumption in these model studies is that the chiral order parameter is constant in space. 
Allowing for spatially varying order parameters, it turns out that there can be a region in the phase diagram where 
an inhomogeneous phase is favored over homogeneous ones (see Ref.~\cite{Buballa:2014tba} for a review).
In particular,
in the NJL model it was found that the first-order phase boundary is entirely 
covered by an inhomogeneous phase~\cite{Nickel:2009wj}.
In the chiral limit there is then a so-called Lifshitz point (LP) where the three different phases,
i.e., the homogeneous and inhomogeneous chirally broken ones, and the restored phase meet. 
Moreover, it was shown within a Ginzburg-Landau (GL) analysis that the LP exactly coincides
with the TCP obtained when only considering homogeneous phases~\cite{Nickel:2009ke}.
The same behavior, which was already  known from the 1+1 dimensional Gross-Neveu model~\cite{Schnetz:2004vr}
was also found in the QM model for a special choice of the sigma-meson mass~\cite{Carignano:2014jla}. 

Away from the chiral limit, the situation is less clear. 
In Ref.~\cite{Andersen:2018osr} it was reported for the QM model that, when the symmetry breaking parameter is increased, 
the inhomogeneous phase quickly shrinks and disappears already at a pion mass of 37~MeV, way below the physical value. 
For the NJL model, on the other hand, it was found in Ref.~\cite{Nickel:2009wj} that, although the inhomogeneous phase also
shrinks with increasing explicit symmetry breaking, the effect is less dramatic, and the inhomogeneous phase is still 
present for realistic values of the current quark mass $m$.
In particular it was stated that for $m\neq 0$ the inhomogeneous phase still reaches out to the CEP. 
This statement, however, was only based on numerical evidence and not shown with the same rigor as the coincidence of
the LP with the TCP in the chiral limit. 

We therefore want to revisit this question and investigate the relation between the ``tip'' of the inhomogeneous phase 
and the CEP within a GL analysis. 
We restrict ourselves to the NJL model and postpone the investigation of the QM model, which is analogous but
technically more involved, to a later publication.

\section{Ginzburg-Landau analysis of critical and Lifshitz points}

Consider the free-energy density of a physical system at temperature $T$ and chemical potential $\mu$,
given by the value of  the thermodynamic potential 
$\Omega[M]$. The latter is a functional of some order-parameter field
$M(\x)$, which generally depends on the spatial position $\x$. Specifically we may think of $M$ being a space-dependent 
``constituent quark mass''\footnote{{For this discussion we will focus on real order parameters, an assumption which will be justified
in the following section.}} proportional to the chiral condensate $\ave{\bar\psi\psi}$ at $\x$. 
In the case of an exact chiral symmetry we can then expand $\Omega[M]$ about the chirally restored solution $M\equiv 0$
as \cite{Nickel:2009ke}
\beq
       \Omega[M] = \Omega[0] + \frac{1}{V} \int d^3x \Big( \alpha_2 M^2(\x) + \alpha_{4,a} M^4(\x)   + \alpha_{4,b} (\nabla M(\x))^2
       + \dots     
       \Big) ,
\label{eq:Omega_GL_cl}
\eeq
where $V$ is the quantization volume and the GL coefficients $\alpha_i$ are $T$ and $\mu$ dependent functions. 
Odd powers of $M$  are prohibited by chiral symmetry.  
The ellipsis indicates higher-order terms, which we assume to be positive for the stability of the ground state. 

If $\alpha_{4,b}$ is positive, gradients are suppressed and we have a homogeneous ground state, characterized by
$M=\mathit{const}$.
If $\alpha_{4,a}$ is positive as well, the system is in the restored phase ($M=0$) for $\alpha_2>0$ and
in the broken phase for  $\alpha_2<0$ with a second-order phase transition at $\alpha_2=0$.
In contrast,  there is a first-order phase transition for $\alpha_{4,a}<0$.
Hence, the TCP, where the first-order phase boundary goes over into a second-order one, is given by the condition
\beq
       \mathrm{TCP:} \quad \alpha_2 = \alpha_{4,a}=0 \, .
\label{eq:TCP}       
\eeq
For $\alpha_{4,b}<0$, on the other hand, gradients are favored, so that inhomogeneous order parameters become possible.  
Typically the phase transition between the inhomogeneous phase and the restored phase is of second order.
At the LP it meets the second-order phase boundary between the homogeneous broken and restored  phases, 
so that this point is determined by the condition
\beq
       \mathrm{LP:} \quad \alpha_2 = \alpha_{4,b}=0 \, .
\label{eq:LP}
\eeq
We note that away from the LP, $\alpha_{4,b}$ is negative along the second-order phase boundary between inhomogeneous 
and restored phase, balanced by positive contributions from $\alpha_2 > 0$ and higher-order gradient terms.  
As a consequence, while the amplitude of $M$ vanishes at the phase boundary, the wave number of the modulation stays
nonzero and vanishes only at the LP.

\begin{figure}[t]
\centering
\includegraphics[width=.24\textwidth]{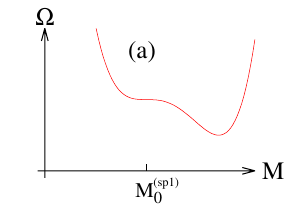}
\includegraphics[width=.24\textwidth]{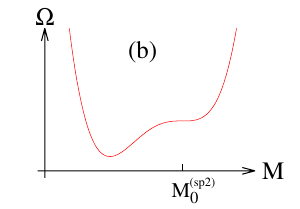}
\includegraphics[width=.24\textwidth]{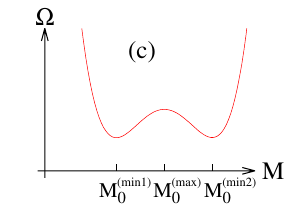}
\includegraphics[width=.24\textwidth]{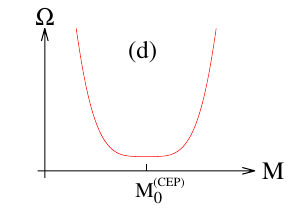}
\caption{ Qualitative behavior of the thermodynamic potential $\Omega$ as a function of a spatially 
homogeneous order parameter $M$
at the left spinodal (a), at the right spinodal (b), at a first-order phase boundary (c), and at the CEP (d). }
\label{fig:Omegaschem}
\end{figure}

The analysis becomes more complicated when the chiral symmetry is explicitly broken, e.g., by the presence of a nonvanishing 
current quark mass. In this case there is no chirally restored solution and we therefore expand the thermodynamic potential
about an --  a priori unknown -- constituent mass $M_0$, 
which we assume to be spatially constant but which may depend on $T$ and $\mu$. 
Writing $M(\x) = M_0 + \delta M(\x)$ and assuming  that $\delta M(\x)$
and its gradients are small, the expansion then takes the form
\beq
       \Omega[M] = \Omega[M_0] + \frac{1}{V} \int d^3x \left(  \alpha_1 \delta M(\x)
        + \alpha_2 \delta M^2(\x) + \alpha_3 \delta M^3(\x) 
       + \alpha_{4,a} \delta M^4(\x)   + \alpha_{4,b} (\nabla \delta M(\x))^2
       + \dots     
       \right) .
\label{eq:Omega_GL_m}
\eeq
Note that here, in contrast to \Eq{eq:Omega_GL_cl}, odd powers of the expansion parameter are allowed. 
In the following we require that $M_0$ corresponds to a stationary point at given $T$ and $\mu$, 
meaning that the linear term of the expansion vanishes, $\alpha_1(T,\mu; M_0) = 0$. 
It is crucial however to keep the cubic term.

Again, we will first consider the case of homogeneous order parameters, $\delta M = \mathit{const}$.
Since there is no restored phase, we cannot have a second-order transition to it,
but it is possible to have a first-order boundary between solutions with different values of $M$, ending at a CEP  
(where the phase transition is second order).
If at temperature $T$ there is a first-order phase transition at a critical chemical potential $\mu_c= \mu_c(T)$,
the thermodynamic potential at $T$ and $\mu_c$ as a function of $M$ has two degenerate minima $M_0^{({\rm min}1)}$ and  $M_0^{({\rm min}2)}$ with $\alpha_2(T,\mu_c;M_0^{({\rm min}1,{\rm min}2)}) > 0$, 
separated by a maximum $M_0^{({\rm max})}$ with $\alpha_2(T,\mu_c;M_0^{({\rm max})}) <0$ (see \Fig{fig:Omegaschem} (c)). 
Then, if we move towards the CEP, the maximum gets more and more shallow until it vanishes at $(T_{\rm{CEP}},\mu_{\rm{CEP}})$, 
corresponding to $\alpha_2(T_{\rm{CEP}},\mu_{\rm{CEP}};M_0^{(\rm{CEP})}) = 0$ (\Fig{fig:Omegaschem} (d)). 
Moreover,  since there is a minimum at this point and not a saddle point, 
$\alpha_3(T_{\rm{CEP}},\mu_{\rm{CEP}};M_0^{(\rm{CEP})})$ vanishes as well.
The CEP is therefore given by the equations
\beq
       \mathrm{CEP:} \quad \alpha_1 = \alpha_2 = \alpha_3=0 \, ,
\eeq
which determine the values of $T_{\rm{CEP}}$ and $\mu_{\rm{CEP}}$ together with the corresponding constituent mass $M_0^{(\rm{CEP})}$. 

An alternative way to derive this condition is to approach the CEP along the spinodal lines. 
Assuming again a first-order phase transition at temperature $T$ and $\mu = \mu_c$, 
the spinodals correspond to the chemical potentials $\mu_{\rm sp1}<\mu_c$ and  $\mu_{\rm sp2}>\mu_c$
at which the left or the right minimum in \Fig{fig:Omegaschem} (c) merges with the maximum to a saddle point
(cf.~\Fig{fig:Omegaschem} (a) and (b)).
If these saddle points are located at $M=M_0^{(\rm sp1)}$ and $M=M_0^{(\rm sp2)}$, respectively,
it follows that $\alpha_2(T,\mu_{\rm sp1};M_0^{(\rm sp1)}) = 0$ and  
$\alpha_3(T,\mu_{\rm sp1};M_0^{(\rm sp1)}) < 0$, while at the right spinodal
$\alpha_2(T,\mu_{\rm sp2};M_0^{(\rm sp2)}) = 0$ 
but  $\alpha_3(T,\mu_{\rm sp2};M_0^{(\rm sp2)}) > 0$.
We thus find again that at the CEP, where the two spinodals meet, both $\alpha_2$ and $\alpha_3$ vanish.

Without a chirally restored phase and a second-order boundary which separates it from the homogeneous 
broken phase, we also cannot have a Lifshitz point,
where this phase boundary is supposed to meet with the boundaries of an inhomogeneous phase. 
It is still possible, however, to have a second-order boundary between an inhomogeneous and a homogeneous phase
along which the amplitude of the {\it oscillating} part of $M(x)$ goes to zero while its wave number in general remains nonzero. 
In analogy to the chiral-limit case we therefore define the pseudo-Lifshitz point (PLP) as the point on this phase boundary
where the wave number vanishes as well.  
It turns out that \Eq{eq:LP} remains basically unchanged for this case, i.e., the condition for the PLP reads
\beq
       \mathrm{PLP:} \quad \alpha_1 = \alpha_2 = \alpha_{4,b}=0 \, .
\label{eq:PLP}
\eeq
This can be shown by a GL analysis of the second-order phase boundary in exactly the same way as done in 
Ref.~\cite{Nickel:2009ke} for the chiral limit. 
A related but perhaps more transparent proof will be given in the next section in the context of a stability analysis 
of the homogeneous phase.

\section{Stability analysis}
\label{sec:stab}

To be specific we consider the standard NJL-model Lagrangian
\beq
       \mathcal{L} = \bar\psi \left( i \slashed{\partial} - m\right)\psi  + G\left\{ (\bar\psi \psi)^2 + (\bar\psi i\gamma_5\vec\tau \psi)^2\right\}
\eeq
for a two-flavor quark field $\psi$ with $N_c = 3$ colors and bare mass $m$, 
and a chirally symmetric four-point interaction in the scalar-isoscalar and pseudoscalar-isovector channels
with coupling constant $G$.  
Allowing for (possibly space dependent but time independent)  scalar and pseudoscalar condensates 
\beq
       \phi_S(\x) = \ave{\bar\psi(\x)\psi(\x)}, \quad \phi_P(\x) = \ave{\bar\psi(\x)i\gamma_5\tau^3\psi(\x)},
\eeq
the mean-field thermodynamic potential at temperature $T$ and quark chemical potential $\mu$ takes the form 
\beq
       \Omega_\mathrm{MF} 
       = -\frac{T}{V} \mathbf{Tr}\, \log\left( \frac{S^{-1}}{T} \right) + G\, \frac{1}{V} \int d^3x \, \left( \phi_S^2(\x) + \phi_P^2(\x) \right),
\eeq
where $V$ is again a quantization volume, and the functional trace runs over the Euclidean space
$V_4 = [0, \frac{1}{T}] \times V$,
Dirac, color, and flavor degrees of freedom. The inverse dressed quark propagator is given by
\beq
       S^{-1}(x) = i\slashed{\partial} + \mu\gamma^0 - m + 2G  \left( \phi_S(\x) + i\gamma_5\tau^3 \phi_P(\x) \right) .
\eeq

We first neglect the possibility of inhomogeneous condensates. At nonvanishing $m$ the ground state at given $T$ and $\mu$
is then given by a constant scalar condensate, $\phi_S(\x) = \phi_{S,0} $, and a vanishing pseudoscalar condensate,
$\phi_P = 0$. Accordingly the inverse quark propagator takes the form
\beq
       S_0^{-1}(x) =  i\slashed{\partial} + \mu\gamma^0 - M_0 \,, \quad  M_0 = m -2G \phi_{S,0} \,,
\label{eq:S0}
\eeq
corresponding to a free particle with constituent quark mass $M_0$.

Next we consider inhomogeneous fluctuations around the homogeneous ground state,
\beq
       \phi_S(\x) = \phi_{S,0} + \delta \phi_S(\x) \,, \quad  \phi_P(\x) \equiv \delta \phi_P(\x) \,.
\label{eq:spfluc}       
\eeq
Searching for a possible second-order phase boundary between the homogeneous and an inhomogeneous phase,
we assume that the amplitudes (but not necessarily the gradients), of the fluctuations are small and expand $ \Omega_\mathrm{MF}$
in powers of $\delta \phi_S$ and $\delta \phi_P$. 
The thermodynamic potential can then be written as
\beq
       \Omega_\mathrm{MF} = \sum_{n=0}^\infty \Omega^{(n)} \,,
\eeq
where $\Omega^{(n)}$ is of the $n$th order in the fluctuating fields.
Specifically we obtain
\beq
       \Omega^{(1)} = \frac{T}{V} \mathbf{Tr} \left( S_0 \delta\hat\Sigma\right) 
       + 2G \phi_{S,0} \, \frac{1}{V} \int d^3x \, \delta \phi_S(\x)  
\eeq
for the linear contribution,
\beq
       \Omega^{(2)} = \frac{1}{2}\frac{T}{V} \mathbf{Tr} \left( S_0 \delta\hat\Sigma\right)^2 
       + G \, \frac{1}{V} \int d^3x \, \left( \delta \phi_S^2(\x) + \delta\phi_P^2(\x) \right)
\eeq
for the quadratic contribution, and 
\beq
       \Omega^{(n\geq3)} = \frac{1}{n}\frac{T}{V} \mathbf{Tr} \left( S_0 \delta\hat\Sigma\right)^n 
\eeq
for all higher-order contributions, where
 $\delta\hat\Sigma(\x) = -2G  \left( \delta\phi_S(\x) + i\gamma_5\tau^3 \delta\phi_P(\x) \right)$
is the quark self-energy related to the fluctuating fields.

Assuming spatially periodic condensates we can perform the Fourier decompositions
\beq
       \delta\phi_{S}(\x) = \sum_{\q_k} \delta\phi_{S, \q_k} \,e^{i\q_k\cdot \x}\,, \quad
       \delta\phi_{P}(\x) = \sum_{\q_k} \delta\phi_{P, \q_k} \,e^{i\q_k\cdot \x} \,,
\label{eq:Fourier}
\eeq
with $\q_k$ being the elements of the corresponding reciprocal lattice.
We require that the condensates and, thus, their fluctuations are real functions in coordinate space.
The Fourier coefficients then obey the relations $\delta\phi_{S, -\q_k} = \delta\phi_{S, \q_k}^*$ and
$\delta\phi_{P, -\q_k} = \delta\phi_{P, \q_k}^*$.
Recalling that $S_0$ is the standard propagator of a free fermion with constant mass $M_0$ it is then straightforward
to evaluate the above expressions for $\Omega^{(n)}$. 
For the linear contribution we find
\beq
       \Omega^{(1)} = -\delta\phi_{S,\mathbf{0}} \left(M_0 - m + 2G M_0\, F_1\right)
\label{eq:Omega1q}
\eeq
where for later convenience we define 
\beq
       F_n = 8N_c \int \frac{d^3p}{(2\pi)^3}\, s_n(\p)
\label{eq:Fn}
\eeq
with
\beq       
       s_n(\p) = T\sum\limits_j \frac{1}{[(i\omega_j +\mu)^2 - \p^2 -M_0^2]^{n}} 
\eeq
and fermionic Matsubara frequencies $\omega_j = (2j+1)\pi T$.

\Eq{eq:Omega1q} shows that only the homogeneous ($\q_k = \mathbf{0}$) part of the scalar fluctuations contribute to this order.
However, since we assumed $\phi_{S,0}$ to be the homogeneous ground-state solution, it follows that  $\Omega^{(1)}$
should vanish. In turn this means that the term in parentheses must be equal to zero, which is nothing but the gap equation
for $M_0$.

For the quadratic contribution one finds
\beq
        \Omega^{(2)} = 2G^2 \sum\limits_{\q_k} \left\{ \left| \delta\phi_{S,\q_k}\right|^2 \, \Gamma_S^{-1}(\q_k^2)
                                                                              + \left| \delta\phi_{P,\q_k}\right|^2 \, \Gamma_P^{-1}(\q_k^2) \right\} \,,
\label{eq:Omega2q}
\eeq
with the inverse scalar and pseudoscalar correlation functions 
\begin{alignat}{1}
\Gamma_S^{-1}(\q^2) &= \frac{m}{M_0} \frac{1}{2G}  - \frac{1}{2} \left(\q^2 + 4M_0^2\right) L_2(\q^2)\,,
\label{eq:GammaS}
\\
\Gamma_P^{-1}(\q^2) &= \frac{m}{M_0} \frac{1}{2G}  - \frac{1}{2} \q^2  L_2(\q^2)\,,
\label{eq:GammaP}
\end{alignat}
which can be interpreted as inverse meson propagators at vanishing energy and nonzero 3-momentum $\q$. 
Here the gap equation has been exploited to eliminate a term proportional to $F_1$, while
\beq
        L_2(\q^2) = -8N_c \int \frac{d^3p}{(2\pi)^3}\, 
        T\sum\limits_n \frac{1}{[(i\omega_n +\mu)^2 - (\p+\q)^2 -M_0^2][(i\omega_n +\mu)^2 - \p^2 -M_0^2]}\,.
\eeq

\begin{figure}[t]
\centering
\includegraphics[width=.42\textwidth]{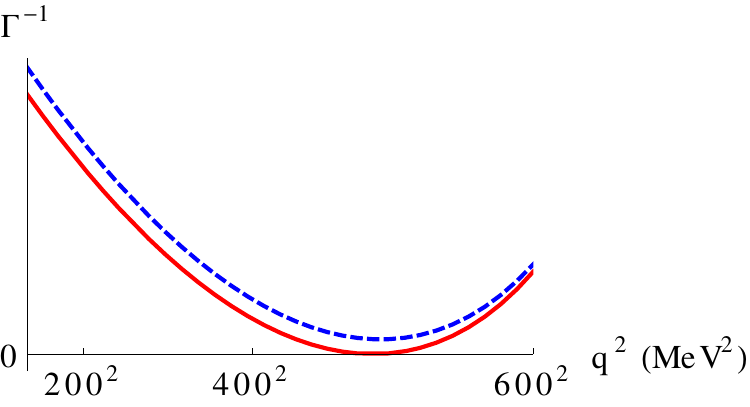}
\caption{Inverse correlation functions $\Gamma^{-1}_S$ (red solid line) and $\Gamma^{-1}_P$ (blue dashed line)
for $m=10$~MeV, $T= 10$~MeV and $\mu = 344$~MeV as functions of the momentum $\q^2$.}
\label{fig:Gamma}
\end{figure}

As obvious from \Eq{eq:Omega2q} the homogeneous ground state is unstable against the formation of inhomogeneous modes
if $\Gamma_S^{-1}$ or $\Gamma_P^{-1}$ is negative in some region of $\q$. 
According to \Eqs{eq:GammaS} and (\ref{eq:GammaP}) a necessary condition for this to happen is that the function $L_2(\q^2)$ 
is positive in this regime (since $m$, $M_0$ and $G$ are positive). 
In this case we always have $\Gamma_S^{-1} < \Gamma_P^{-1}$, meaning that the instability
occurs first in the scalar channel.
This is illustrated in \Fig{fig:Gamma} where $\Gamma^{-1}_S$ and $\Gamma^{-1}_p$ 
at $T= 10$~MeV and $\mu = 344$~MeV are displayed as functions of the momentum $\q^2$ for 
a current quark mass $m=10$~MeV, roughly corresponding to a vacuum pion mass $m_\pi = 135$~MeV.\footnote{
In all our numerical calculations we use Pauli-Villars regularization with three regulator terms and adopt the 
coupling constant $G$ and the Pauli-Villars parameter $\Lambda$ from Ref.~\cite{Nickel:2009wj} where they
have been fitted in the chiral limit  to a vacuum constituent quark mass of 300~MeV and a pion decay constant of 88~MeV.}
While the function $\Gamma^{-1}_S$ just touches the zero-axis, indicating that the chosen values of $T$ and $\mu$ 
correspond to a point on the phase boundary, $\Gamma^{-1}_P$ is strictly positive, i.e., the system is 
stable against pseudoscalar fluctuations.
This is qualitatively different from the situation in the restored phase of the chiral limit, 
which was discussed in Ref.~\cite{Nakano:2004cd}. There we have $M_0 = 0$ 
and thus the instabilities in the scalar and pseudoscalar channels occur simultaneously.

\begin{figure}[t]
\centering
\includegraphics[width=.48\textwidth]{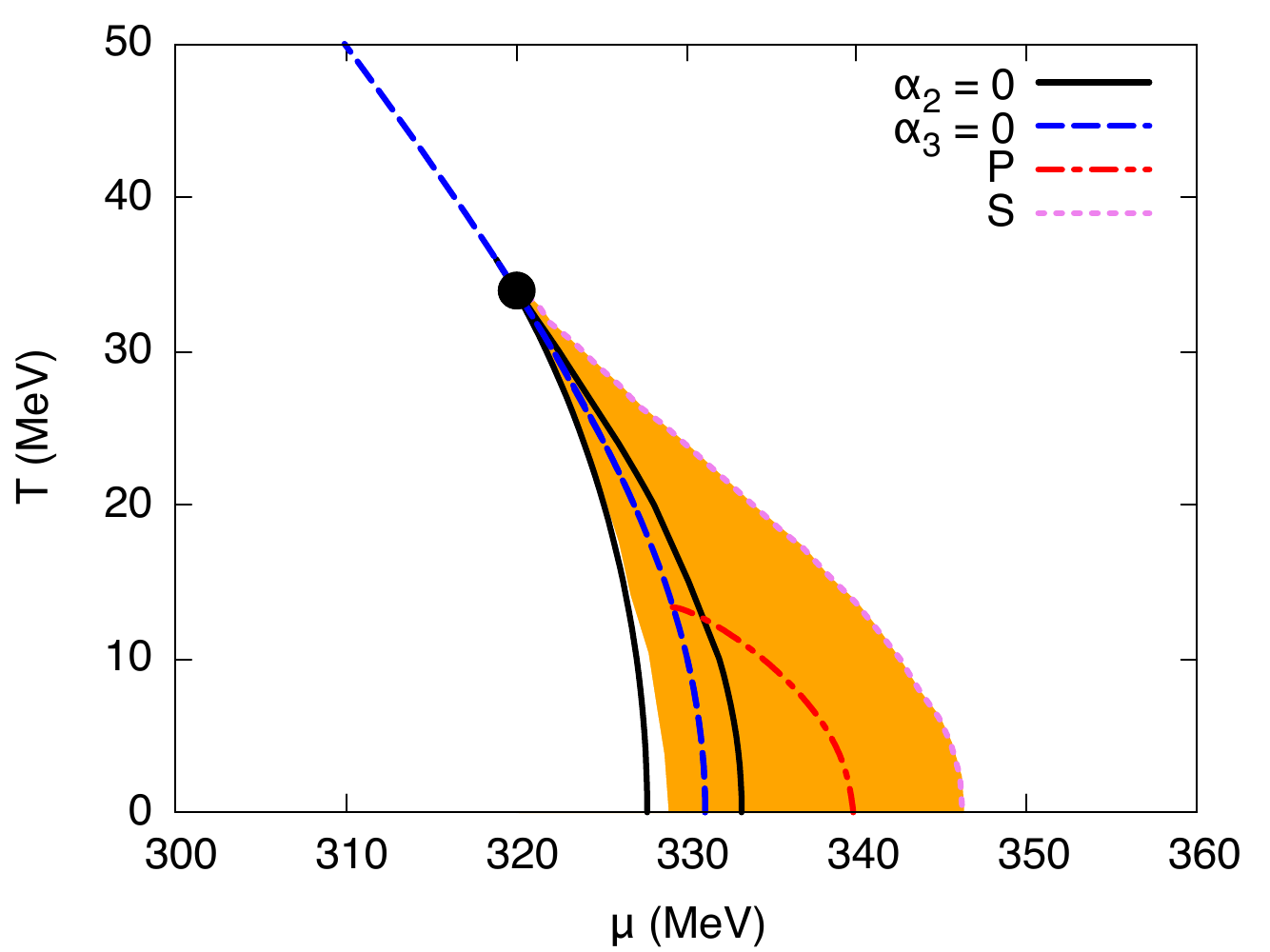}
\caption{
Relevant lines in the phase diagram of the NJL model for $m = 10$~MeV: 
GL coefficients $\alpha_2 = 0$ (black solid) and  $\alpha_3 = 0$ (blue dashed), meeting at the CEP (black dot).
As discussed in Sec.~\ref{sec:GLc}, the  $\alpha_3 = 0$ line is identical to the  $\alpha_{4,b} = 0$ line and
therefore the CEP coincides with the PLP.
At each point $M_0$ was determined by simultaneously solving the equation $\alpha_1 = 0$. 
The shaded area indicates the region where the inhomogeneous solution of Ref.~\cite{Nickel:2009wj} is favored 
over homogeneous phases. 
Its phase boundary to the right coincides with the instability line against scalar fluctuations (pink dotted).
The red dash-dotted line indicates the line where the instability against pseudoscalar fluctuations would occur if 
scalar fluctuations were suppressed.
} 
\label{fig:pd}
\end{figure}

In particular we find that near the second-order phase boundary the inhomogeneous mode is purely scalar for $m\neq 0$,
so that modulations like the dual chiral density wave \cite{Nakano:2004cd} or modifications thereof which contain 
pseudoscalar condensates \cite{Karasawa:2013zsa,Takeda:2018ldi}
are excluded in this regime.\footnote{The picture might change if magnetic fields  \cite{Frolov:2010,Abuki:2018iqp}
or vector interactions \cite{Carignano:2018hvn}
are included.}
This can also be seen in \Fig{fig:pd}. 
The shaded area corresponds to the inhomogeneous region of Ref.~\cite{Nickel:2009wj}  where a so-called
real kink crystal (``solitonic'') modulation was considered, which is purely scalar and allows for a smooth transition 
to the homogeneous phase, consistent with our ansatz \Eq{eq:spfluc}. 
Its phase boundary on the right coincides with the border of the instability region 
with respect to scalar fluctuations (pink dotted line). 
If instead only pseudoscalar fluctuations were allowed, the inhomogeneous phase would already end at the 
red dash-dotted line, i.e., at considerably lower chemical potentials.\footnote{Note that 
this type of analysis cannot say anything about the left boundary of the inhomogeneous phase where the amplitudes of the oscillating fields are large.}

In the following we will therefore drop the pseudoscalar condensate and concentrate on the scalar channel.  
In analogy to the definition of $M_0$ in \Eq{eq:S0} we then define $\delta M(\x) = -2G\delta\phi_S(\x)$,
which can be identified with the $\delta M(\x)$ introduced in the previous section. 	
Plugging this, together with the Fourier decomposition \Eq{eq:Fourier}, into \Eq{eq:Omega_GL_m} and comparing
the terms proportional to $| \delta\phi_{S,\q_k}|^2$ with \Eq{eq:Omega2q} we can identify
\footnote{Generally this yields the coefficients associated with terms of the form $(\nabla_i\nabla_j\dots M)^{2}$,
which are used in the improved GL analysis developed in \cite{Carignano:2017meb}.
}
\beq
       \alpha_2 = \frac{1}{2} \Gamma_S^{-1}(0) \,, \quad
       \alpha_{4,b} = \frac{1}{2}\left. \frac{d\Gamma_S^{-1}}{d\q^2}\right|_{\q^2=0} \;.     
\label{eq:alphaL2}
\eeq

As illustrated in \Fig{fig:Gamma},
the second-order phase boundary between homogeneous and inhomogeneous phase is given by the condition
that $\Gamma_s^{-1}(\q^2)$ just touches the $\Gamma_s^{-1}=0$-axis at a single momentum.
This momentum is then the wave number of the inhomogeneous modulation at the phase boundary.
Hence, at the PLP, which we defined to be the point of the phase boundary where the wave number vanishes,
we have $\Gamma_S^{-1}(0) = 0$ and $\frac{d\Gamma_S^{-1}}{d\q^2}|_{\q^2=0}= 0$.
Comparing this with \Eq{eq:alphaL2}, we conclude that $\alpha_2 =  \alpha_{4,b} = 0$ at the PLP, 
in agreement with \Eq{eq:PLP}.

Of course, what we are really interested in is the tip of the inhomogeneous phase in the phase
diagram, which, from a logical point of view, is not necessarily the same as our definition of the PLP.
However, if one continues to move in the direction of the phase boundary beyond the PLP, 
the value of $\q^2$ where $\Gamma_S^{-1}$  touches the zero-axis becomes negative, 
and therefore there is no instability in this region. 
Here we assumed that $\Gamma_S^{-1}$ is a smooth function of $T$ and $\mu$ at the PLP.
But even if this is not the case, e.g., if there is  a discontinuous jump of $M_0$,
this should reflect itself in the behavior of the phase boundary at this point,
most likely producing a kink.
So in any case the PLP, as we defined it, should be a significant point of the phase boundary. 
Indeed, numerically we confirm that it corresponds to the tip of the inhomogeneous phase.

\section{Evaluation of the Ginzburg-Landau coefficients}
\label{sec:GLc}

According to \Eq{eq:alphaL2}, we can calculate $\alpha_2$ and $\alpha_{4,b}$ directly from $\Gamma_s^{-1}$. 
Analogously, $\alpha_1$ can be derived from  $\Omega^{(1)}$, while $\alpha_3$ and $\alpha_{4,a}$ can be obtained
from $\Omega^{(3)}$ and $\Omega^{(4)}$, respectively. 
Instead of using the Fourier decomposition of the condensates, as done in the previous section, an equivalent but 
perhaps more straightforward procedure to calculate the Ginzburg-Landau coefficients is to keep the coordinate-space 
representation and perform a gradient expansion \cite{Nickel:2009ke}. One finds
\begin{alignat}{1}
\alpha_1 & = \frac{M_0-m}{2G} + M_0 F_1 \,,
\\
\alpha_2 & = \frac{1}{4G} + \frac{1}{2} F_1 +  M_0^2 F_2 \,,
\\
\alpha_3 & = M_0 \left( F_2 +  \frac{4}{3} M_0^2 F_3 \right) \,,
\\
\alpha_{4,a} & = \frac{1}{4} F_2 +  2  M_0^2 F_3 + 2 M_0^4 F_4 \,,
\\
\alpha_{4,b} & = \frac{1}{4} F_2 +  \frac{1}{3}  M_0^2 F_3 \,,
\label{eq:alpha4bres}
\end{alignat}
with the integrals $F_n$ defined in \Eq{eq:Fn}.
We immediately see that the stationarity condition $\alpha_1 = 0$ is equivalent to $\Omega^{(1)} = 0$, \Eq{eq:Omega1q},
as expected. In addition, the above results have several interesting consequences:
\begin{itemize}

\item[(i)] In the chiral limit, where we can expand about the restored solution $M_0 = 0$, we reproduce that
              $\alpha_{4,a} = \alpha_{4,b}$~\cite{Nickel:2009ke}, meaning that TCP and LP coincide.         

\item[(ii)] Assuming $M_0 \neq 0$ and then taking the limit $M_0 \rightarrow 0$, the $\alpha_3 = 0$ line  
              converges to the $F_2=0$ line, which also determines the $\alpha_{4,a} = 0$ line in this limit.
              Hence the CEP converges to the TCP in the chiral limit, as one would expect. 

\item[(iii)] For arbitrary values of $M_0 \neq 0$ we find  $\alpha_3 = 4M_0\,\alpha_{4,b}$ and, thus, the PLP 
               coincides with the CEP. 

\end{itemize}

The lines $\alpha_2 = 0$ and $\alpha_3 = 0$ (coinciding with $\alpha_{4,b} = 0$) are also shown in \Fig{fig:pd}.
As one can see, both spinodals (i.e., the left and the right part of the $\alpha_2 = 0$ line)
and the $\alpha_3 = 0$ line become parallel at the CEP.
Therefore from a practical point of view the $\alpha_3 = 0$ line does not really help to localize the CEP, which is
found more easily by determining the point where the two spinodals meet. 
However, as we have seen above, the condition $\alpha_3 = 0$ at the CEP is very important conceptually in order to 
show its coincidence with the PLP.

In practical calculations we have to deal with the fact that some of the integrals are divergent and have to be 
regularized. One must then be careful not to spoil the above relations by an improper choice of the regularization
procedure.
In fact, the derivation of the Ginzburg-Landau coefficients relies at several places on integrations by parts 
and the assumption that surface terms can be dropped. 
This was already pointed out  in Ref.~\cite{Nickel:2009ke} for the chiral limit and gets additional importance
in our case for showing the proportionality of $\alpha_{4,b}$ to $\alpha_3$. 
Here a straightforward calculation along the lines of  Ref.~\cite{Nickel:2009ke} yields
\beq
       \alpha_{4,b} = \frac{1}{4} F_2  -   M_0^2 \frac{16}{3}N_c \int \frac{d^3p}{(2\pi)^3}\, \p^2 s_4(\p) \,,
\eeq
and the result given in \Eq{eq:alpha4bres} is obtained by noting that $s_4 = \frac{1}{6 |\p|} \frac{\partial s_3}{\partial|\p|}$
and integrating by parts, assuming that the surface terms can be dropped. 
As already found out earlier when dealing with inhomogeneous phases, we should therefore not regularize the integrals 
by a momentum cutoff,  while for instance Pauli-Villars (which was employed in all numerical calculations presented here)
or proper-time regularization are permitted.

\section{Discussion}

The coincidence of the PLP (which we have argued to be the tip of the inhomogeneous phase) with the CEP
in the NJL model with a nonzero current quark mass is our main result. 
It means that there is an inhomogeneous phase in the model if the analysis of homogeneous phases predicts the
existence of a first-order chiral phase transition with a CEP.
Of course this result does not tell us whether there is a CEP in the first place, and how it behaves as a function 
of the quark mass. 

\begin{figure}[t]
\centering
\includegraphics[width=.4\textwidth]{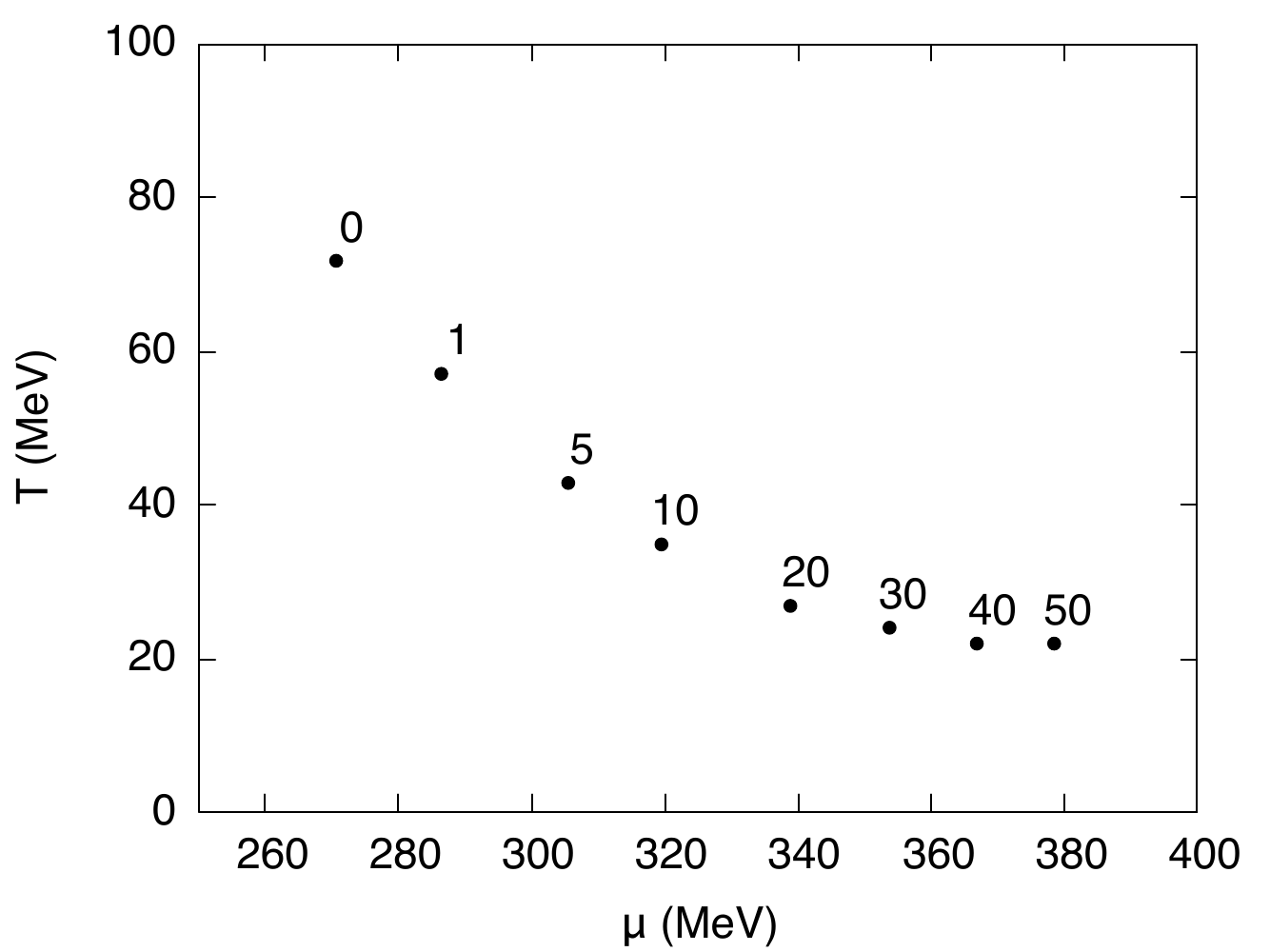}
\caption{Positions of the critical point (TCP$=$LP for $m= 0$, CEP$=$PLP for $m\neq0$)
in the phase diagram for different values of the current quark mass $m$ (as indicated, in MeV).
The relation between $m$ and the vacuum pion mass is $(m/\mathrm{MeV},m_\pi/\mathrm{MeV}) =$ 
$(0,0)$, $(1,43)$, $(5,96)$, $(10,135)$, $(20,191)$, $(30,235)$, $(40,271)$, $(50,303)$.
} 
\label{fig:CPmov}
\end{figure}

In \Fig{fig:CPmov} we therefore show the positions of the CEP (and hence of the PLP) in the phase diagram 
for different values of $m$, keeping the other model parameters as in the numerical examples shown before. 
For small $m$ we find that the temperature of the PLP strongly decreases with increasing quark mass, suggesting 
that the inhomogeneous phase will quickly disappear at higher $m$. 
It turns out, however, that this is not the case, as the movement of the PLP as a
function of $m$ slows down, behaving more like the increase of $m_\pi \propto \sqrt{m}$.
Moreover, the temperature levels off, although we cannot exclude that this is a regularization artifact.\footnote{For
even higher quark masses we find that the temperature of the PLP rises again.}
It is therefore not obvious from these calculations whether or not inhomogeneous condensates could also play a role in the
strange-quark sector. 

Independent of the numerical details, the coincidence of CEP and PLP means that the possibility of having an
inhomogeneous phase should be taken as seriously as the possibility of the existence of a CEP.
We note that the coincidence of the two points could have been anticipated not only from the same feature in the
1+1 dimensional Gross Neveu model \cite{Schnetz:2005ih} or from the numerical evidence of Ref.~\cite{Nickel:2009wj} 
but already from the earlier observation that near the CEP the scalar meson propagator has a space-like peak 
which at the CEP diverges for $|\q| \rightarrow 0$~\cite{Fujii:2003bz}.
As discussed in Sec.~\ref{sec:stab}, this indicates an instability towards an  ``inhomogeneous'' mode
with vanishing wave number at this point, in agreement with our definition of the PLP.  

We should mention that already in the chiral limit the coincidence of TCP and LP does not necessarily hold 
if additional interaction terms are considered. However, as shown in Ref.~\cite{CNB:2010} 
for the case of vector interaction, these terms typically affect the TCP stronger than the LP, so that the 
inhomogeneous phase is rather robust under model extensions. We therefore expect a similar behavior at nonzero $m$.

Finally, we recall that fluctuations beyond the mean-field approximation discussed here are expected to play an important role on the phase structure of the model, as they are known to prevent the formation of any true one-dimensional long-range ordering at finite temperature \cite{Baym:1982}. One expects instead a phase 
characterized by a quasi-long-range ordering, similar to that of smectic liquid crystals.
Moreover, it was recently argued that, at least in the chiral limit, the presence of fluctuations prevents the coefficients 
$\alpha_2$ and $\alpha_{4,b}$ from vanishing simultaneously, so that a LP cannot exist and gets instead replaced by a ÒLifshitz regimeÓ in the neighboring region of the phase diagram \cite{Pisarski:2018bct}.  
The properties of low-energy modes and their effects  on crystalline chiral condensates in the chiral limit have been explicitly investigated in \cite{Lee:2015bva} and \cite{Hidaka:2015xza} by employing GL functionals, and more recently in \cite{Yoshiike:2017kbx}. It would be very interesting to extend these analyses to finite quark masses, as the explicit breaking of chiral symmetry might affect the properties of these fluctuations and the possible disappearance of the LP from the phase diagram \cite{Hidaka:2015xza}. The GL functional employed in this work could provide an effective starting point for this kind of analysis.

It is rather straightforward to extend our analysis to the QM model with an explicitly symmetry-breaking term. 
A calculation analogous to  Sec.~\ref{sec:stab} then reveals that instabilities towards inhomogeneous modes 
are again signalled by poles in the dressed scalar and pseudoscalar propagators at vanishing energy and finite 
momentum. Unlike in the NJL model, however, these poles are not only determined by the quark-antiquark 
polarization functions (which again favor instabilities in the scalar modes) but also by the tree-level
meson masses. The situation is therefore more involved and will be discussed in a future publication.\footnote{A 
stability analysis in the pseudoscalar channel has been performed in Ref.~\cite{Tripolt:2017zgc}
within a functional renormalization-group framework.} 
In particular it will be interesting to see whether the early disappearance of the inhomogeneous phase 
at nonzero pion mass, which has been found in Ref.~\cite{Andersen:2018osr} for a dual chiral density wave
modulation, will still persist if a purely scalar modulation, like the real kink crystal, is considered. 

Ultimately, we are of course interested in the phase diagram of QCD. In Ref.~\cite{Muller:2013tya} inhomogeneous phases
have been studied within a Dyson-Schwinger approach applied to two-flavor QCD in the chiral limit.
Using a relatively simple truncation scheme, an inhomogeneous selfconsistent  solution of the Dyson-Schwinger 
equations was found to exist in a region of the phase diagram quite similar to that in the  NJL model. 
In particular, within numerical precision, the LP agrees with the TCP in that approach. Our present finding of the
coincidence of PLP and CEP might therefore also have relevance for QCD away from the chiral limit.

\subsection*{Acknowledgments}
We acknowledge support by the Deutsche Forschungsgemeinschaft (DFG, German Research Foundation) through the CRC-TR 211 `Strong-interaction matter under extreme conditions' - project number 315477589 - TRR 211.
S.C. also acknowledges financial support by the Fondazione Angelo Della Riccia.

\bibliography{biblio1008}

\end{document}